\documentclass{emulateapj}
\usepackage{natbib}
\usepackage{epsfig}
\begin{document}


\title{Very High Energy $\gamma$-ray Afterglow Emission of Nearby Gamma-ray Bursts}


\author{R. R. Xue$^{1,2}$, P. H. Tam$^{3}$, S. J. Wagner$^{3}$, B. Behera$^{3}$, Y. Z. Fan$^{4,1}$ and D. M. Wei$^{1,5}$}
\affil {$^1$ Purple Mountain Observatory, Chinese Academy of
Sciences,
Nanjing 210008, China.\\
$^2$ Graduate School, Chinese Academy of Sciences,
Beijing, 100012, China.\\
$^3$ Landessternwarte, Universit\"{a}t Heidelberg, K\"{o}nigstuhl, Germany.\\
$^4$ Niels Bohr International Academy, Niels Bohr Institute,
University of Copenhagen, Blegdamsvej 17, DK-2100 Copenhagen,
Denmark.\\
 $^5$ Joint Center for Particle Nuclear Physics
and Cosmology of Purple Mountain Observatory -- Nanjing University,
Nanjing 210008, China. }\email{rrxue@pmo.ac.cn (RRX) ;\\
phtam@lsw.uni-heidelberg.de (PHT)}

\begin{abstract}
The synchrotron self-Compton (SSC) emission from Gamma-ray Burst
(GRB) forward shock can extend to the very-high-energy (VHE;
$E_\gamma
> $100~GeV) range. Such high energy photons are rare and are attenuated by the cosmic infrared background
before reaching us. In this work, we discuss the prospect to detect
these VHE photons using the current ground-based Cherenkov
detectors. Our calculated results are consistent with the upper
limits obtained with several Cherenkov detectors for GRB~030329,
GRB~050509B, and GRB~060505 during the afterglow phase. For 5 bursts
in our nearby GRB sample (except for GRB~030329), current
ground-based Cherenkov detectors would not be expected to detect the
modeled VHE signal. Only for those very bright and nearby bursts
like GRB~030329, detection of VHE photons is possible under
favorable observing conditions and a delayed observation time of
$\la$10~hours.
\end{abstract}


\keywords{Gamma Rays: bursts --- ISM: jets and outflows --- radiation
mechanism: non-thermal}

\section{Introduction}
On February 28, 1997, the first X-ray afterglow of a Gamma-ray
burst (GRB) was detected, leading to the identification of its
progenitor at cosmological distances~\citep{costa97}. In a few
days, the afterglow faded away with time as a power law. This
behavior is satisfactorily explained in the spherical (isotropic)
fireball model involving relativistic ejecta decelerated by
circumburst medium~\citep{meszaros97}. The introduction of
collimated jets relaxes the energy requirement 
of
GRBs by a factor of several hundred, as well as explains
the steeper temporal decay of afterglows ~\citep{rhoads99,sari99}.

GRBs are extra-galactic sources of GeV and probably higher energy
photons. Evidences of distinct high-energy component have been
accumulated by EGRET onboard the Compton Gamma-Ray Observatory:
(1) \citet{hurley94} reported the detection of long-duration
MeV--GeV emission of GRB~940217 lasting up to 1.5 hour after the
keV burst, including an $\sim$18 GeV photon. This burst is the
longest and the most energetic among those GRBs with detected
high-energy emission so far; (2) \citet{Gonzalez03} revealed a
high-energy component of GRB~941017 temporally and spectrally
different from the low-energy component.

In the fireball model, synchrotron emission of shock-accelerated
electrons is commonly thought to produce prompt $\gamma$-ray
emission as well as afterglow emission at lower energies
\citep[e.g.,][]{sari98}. It is natural to expect that these
photons are inverse-Compton up-scattered by electrons, giving rise
to a higher energy component peaking at sub-GeV to TeV
energies~\citep{wei98,sari01}. When electrons scatter the
self-emitting synchrotron photons, synchrotron self-Compton (SSC)
emission is resulted. In the external shock scenario, the temporal
profile of the SSC emission from forward shock electrons is
similar to that of the low energy afterglow emission and no
significant time lag is expected.

The \emph{Swift} satellite, thanks to its rapid response time and
accurate localization, has started a new era of research on GRBs.
Different modifications to the standard afterglow model are put
forward to explain the peculiar behaviors exhibited in the X-ray
light curves, in particular the shallow declining
phase~\citep{zhang06,nousek06}. Recently, the SSC emission of the
modified forward shock has been extensively discussed in the
literature~\citep{wei07,gou07,Fan08_mn,galli07,yu07} and applied
to the case of GRB~940217~\citep{wei07}.

The AGILE (Astro-rivelatore Gamma a Immagini L'Eggero) satellite,
launched on April 23, 2007, is dedicated to high-energy $\gamma$-ray
astronomy. The Fermi Gamma-ray Space Telescope (FGST) was launched
on June~11, 2008. The Large Area Telescope (LAT) onboard covers the
energy range from 20 MeV to 300 GeV and its effective area is about
5 times larger than that of EGRET at GeV energies. The first GRB
observations with LAT have resulted in detection of photons with
energies larger than $\sim1$~GeV from several
GRBs~\citep{Omodei08_GCN8407,Tajima08_GCN8246}. \citet{dermer00},
\citet{zhang01b}, and \citet{wang01} predicted promising and
detectable SSC emission from the forward shock with FGST out to
z$\sim$1.

Most of the discussions in the literature have focused on the
afterglow emission from tens of MeV to GeV. LAT can also detect
very-high-energy (VHE; $>$100~GeV) afterglow emission. However,
with a small effective area $\sim 10^4~{\rm cm^2}$, it is very
hard to have a significant detection at such high energy. Imaging
atmospheric Cherenkov telescopes  such as
H.E.S.S.\footnote{http://www.mpi-hd.mpg.de/hfm/H.E.S.S./H.E.S.S..html},
MAGIC\footnote{http://wwwmagic.mppmu.mpg.de/}, and
VERITAS\footnote{http://veritas.sao.arizona.edu/} may serve better
at energies above $\sim$100~GeV because of their much larger
effective area ($\sim 10^8-10^9~{\rm cm^2}$) and a high rejection
rate of hadronic background. The effective collecting area of Cherenkov telescopes
increases with energy~\citep{aharonian06b}. Some of these large area Cherenkov
detectors have been used to set constraints on the possible VHE
afterglow component of GRBs~\citep{albert07,horan07,aharonian09}.
It is thus desirable to see whether these results are consistent
with the predictions of the fireball model. Our aim of this paper
is also to investigate the prospect of significant detections in
the future. To have a reliable estimate of the afterglow emission
at energies above 100 GeV, one needs to calculate the forward
shock emission (both synchrotron and SSC emission of the shocked
electrons) carefully. The attenuation of VHE photons by the cosmic
infrared background is also taken into account.Since the
attenuation effect  for photons with an energy $>$100 GeV is more
severe for high-redshift GRBs, we limit our GRB sample to nearby
events.

This paper is organized as follows: in
\S\ref{sect_afterglow_modeling}, we describe the GRB afterglow
model, introduce the code that is used in the afterglow modeling
and the calculation of the SSC emission from GRB forward shock. In
\S\ref{sect_model_prediction}, we present the expected results of
the SSC model using reasonable parameter values for GRBs. In
\S\ref{sect_vheafterglow_fromnearbygrbs}, we describe the GRB
sample which includes six nearby GRBs with sufficient
multi-wavelength afterglow data and predict their corrected energy
flux after the attenuation by the cosmic background during the
afterglow phase, which is then compared with the available
observational data. We summarize our results and discuss their
implications in \S\ref{sect_discussion}. We conclude in
\S\ref{sect_conclusions}.

\section{afterglow modeling}
\label{sect_afterglow_modeling}
\subsection{GRB Afterglow Model}
\label{subsect_grb_afterglow_model}

While synchrotron emission is widely considered to be responsible
for the radio, optical, and X-ray
afterglows~\citep[e.g.][]{sari98}, inverse Compton scattering  of
forward shock photons is considered in details
by~\citet{wei98,wei00} and \citet{sari01}. Inverse Compton
scattering may considerably change the temporal and spectral
behavior of GRB afterglows, and its cooling effect on electrons
accelerated in external shocks will contribute to the photon
spectra at sub-GeV to TeV energy
range~\citep{meszaros94,dermer00,zhang01b,wang01}.

In the afterglow model, both synchrotron emission and inverse
Compton emission are taken into account. It is assumed that: (1)
the external medium is homogenous with a density $n$ or a wind
profile $n\propto R^{-2}$; (2) the relativistic jet is uniform,
i.e. energy per solid angle is independent of direction within the
jet; (3) the shock parameters ($\epsilon_e$ and $\epsilon_B$,
fractions of the shock energy given to the electrons and the
magnetic field, respectively) are constant; (4) the energy
distribution of electrons accelerated in shocks follows
$dN_\mathrm{e}/dE \propto E^{-p}$; (5) the possible achromatic
flattening in the afterglow light curve is due to energy injection
in the form $E_{k}\propto t^{1-q}$~\citep{cohen99,zhang01a} or
$E_k \propto [1+(t/T)^2]^{-1}$ with $T$ being the initial
spin-down time scale~\citep{dai98}.

The parameters involved in this afterglow model include: $E_{0}$
(the initial isotropic outflow energy), $\theta_{0}$ (the initial
half-angle of the jet), $n$ (the density of the homogeneous
external medium) or $A$ (the wind parameter), $p$ (the power-law
index of energy distribution of shock-accelerated electrons),
$\epsilon_{\rm e}$, and $\epsilon_{\rm B}$ (shock parameters). In
the case where energy injection is necessary, three additional
parameters: $L_{\rm eje}$ (the injected luminosity in the rest
frame), the timescale of energy injection and $q$, are included.

\subsection{A Brief Description of the SSC Model}
The code used in our afterglow modeling and the prediction of the
SSC emission is that developed by~\citet{Fan08_mn}, who carry out
numerical calculations of synchrotron and SSC emission of the
external forward shock in the afterglow phase\footnote{ The cooling
of the forward shock electrons by the so-called ``central engine
afterglow" emission (the emission powered by the re-activity of the
central engine in the afterglow phase, like the flares
\citep[e.g.,][]{nousek06} or the plateaus followed by a sudden drop
\citep[e.g.,][]{zhang09}) and the corresponding inverse Compton
emission can also be calculated self-consistently. But for our
current sample people did not see such emission.}. The reverse shock
emission, predicted in the fireball model but not detected in most
events \citep{roming06,Rykoff09}, is not taken into account.

The key treatments~\citep[see \S3 of][for details]{Fan08_mn} are as
follows: (i) The dynamical evolution of the outflow is followed
using the formulae in~\citet{huang00}, which describes the
hydrodynamics in both relativistic and non-relativistic phases. (ii)
The arbitrary assumption that  the distribution of shocked electrons
is always in a quasi-stationary state is considered to be
unsatisfactory and the energy distribution of electrons is
calculated by solving the continuity equation with the power-law
source function Q = K$\gamma_{e}^{-p}$, normalized by a local
injection rate~\citep{moderski00}. (iii) The observed flux is
integrated over the ``equal-arrival surface". (iv) The Klein-Nishina
correction of the inverse Compton emission has been included. (v)
Energy injection into the outflow is considered necessary in
reproducing some multi-frequency afterglow data of some GRBs. This
may change the dynamics significantly.

By fitting the low-energy multi-waveband afterglows, from radio to
X-ray band, parameters involved in the afterglow model are gotten.
{\it Simultaneous afterglow data in at least two well-separated
wavebands are needed to get a relatively well-constrained set of
parameters.}

A rough estimate of the energy-integrated VHE afterglow flux
(without correction by the cosmic background) is given by
\begin{equation}
F_{>100\rm GeV}\propto \frac{(1+z)L_{\rm ssc}}{D_L^2}
\max\{(\nu_c^{\rm ssc})^{p-2\over 2},~(\nu_m^{\rm ssc})^{p-2\over
2}\}
\end{equation}
where $L_{\rm ssc}$ is the total luminosity of the SSC
emission~\citep[see eq.(23-27) in] [for the expression]{Fan08_mn},
$\nu_m^{\rm ssc}$ and $\nu_c^{\rm ssc}$ are the typical SSC
emission frequency and the SSC cooling frequency of the forward
shock electrons~\citep[see eq.(33-34) in][the case of $k=0$, for
the expressions]{Fan08_mn} and $D_L$ is the luminosity distance of
the event.

\section{model prediction}
\label{sect_model_prediction}  High energy photons, especially those
in the TeV range, will be
attenuated by the cosmic background light. 
Various models of the spectral energy distribution of the cosmic
infrared background are
proposed~\citep{primack01,totani02,kneiske02,stecker06}, but all
these models give comparable opacities for low redshifts. In this
work, a level consistent with a study of two distant blazars and
galaxy counts is used \citep[P0.45,][]{aharonian06a}.

We adopt reasonable values of parameters for nearby GRBs  and
predict the spectra in high-energy to VHE range. After corrected
for the attenuation by extragalactic background, we compare them
with the sensitivity levels of $\gamma$-ray instruments.

\begin{figure}
\begin{center}
\includegraphics[width=280pt]{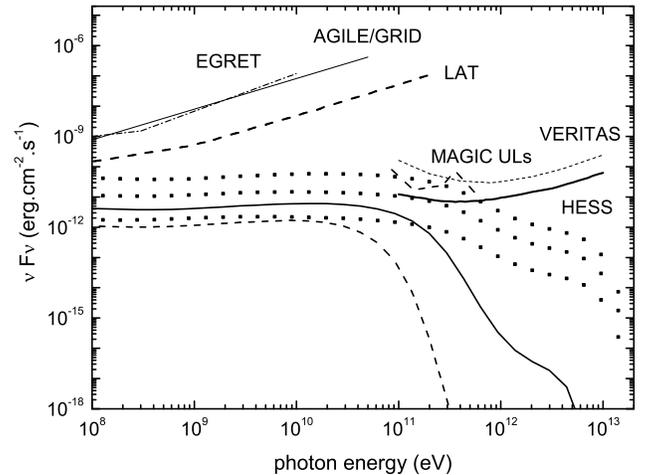}
\end{center}
\caption{Temporal evolution of the observed high energy spectrum of
SSC afterglows. The three dotted lines are the spectra at different
epochs. All spectra, starting from (top) 0.5 hour, 2 hours, and
(bottom) 10~hours, are integrated over 0.5 hour. They are calculated
with the following parameter values: $E_{0}$ =$5\times10^{51}$$ \rm
erg$, $\theta_{0}= 0.4$, $n=1.0 \rm cm^{-3}$, $p=2.2$,
$\epsilon_{\rm e}=0.3$, $\epsilon_{\rm B}=0.01$ and $z=0.16$. The
solid and dashed line are calculated with the same parameters
represented above during the first epoch but occurring at larger
redshift, $z = 0.5$ and $z = 1.0$, respectively. The sensitivity
curves of EGRET~\citep{thompson}, AGILE/GRID~\citep{galli08},
FGST/LAT~\citep{galli08}, VERITAS ~\citep{horan07} and HESS
\citep[assuming $\Gamma=2.6$,][]{Tam09_phdthesis} for an integration
time of 0.5 hour are plotted as labeled. Magic 2-sigma upper limits
derived from 30-min observations of GRB 060206 are also plotted,
taken from Albert et al. (2007). }\label{fig:spectrum}
\end{figure}

Parameters  assumed and the time-averaged spectra, including both
synchrotron and SSC components from the forward shocks, are shown in
Figure~\ref{fig:spectrum}. For this fictitious burst, current
Imaging atmospheric Cherenkov telescopes would be more likely than
space satellites such as EGRET, AGILE/GRID and \emph{FGST}/LAT to
detect the modeled emission, as seen in Figure~\ref{fig:spectrum}.

\section{Very high energy afterglow emission of nearby GRBs}
\label{sect_vheafterglow_fromnearbygrbs} For photons with energy
higher than $\sim 100$~GeV, the attenuation due to interaction
with background photons is significant if the source has a high
redshift. Therefore nearby bursts (those with z$<$0.25) are chosen
in this study.

\subsection{The GRB Sample}
To predict the VHE afterglow emission of  nearby GRBs and compare
the calculated results with sensitivity of different $\gamma$-ray
telescopes, GRBs in our sample must satisfy: nearby events to
alleviate the attenuation effect; at least two independent waveband
low-energy afterglows are recorded to get relatively constrained
parameters; the low-energy afterglow can be reproduced according to
the afterglow model.

In this work, we consider nearby GRBs ($z < 0.25$) 
with relatively high luminosity and multi-wavelength
afterglow data sufficient to meaningfully constrain the properties
of the GRBs (i.e. the model parameter values as described in
\S\ref{subsect_grb_afterglow_model}) up to March 2007. Five GRBs
meet such criteria: GRB~030329, GRB~050509B, GRB~050709,
GRB~060505, and GRB~060614. Though having a
relatively large redshift of z$\sim$0.55, GRB~051221A is also
considered in this work because it is one of the brightest short
GRBs detected so far.

GRB~030329 triggered the High Energy Transient Explorer,
HETE-2~\citep{vanderspek04}.  Very detailed BVRI afterglow light
curves, spanning from $\sim$0.05 to $\sim$ 80 days, were compiled
by~\citet{lipkin04}. \citet{tiengo04} reported XMM-Newton and
Rossi-XTE late-time observations of this burst. Based on the
emission and absorption lines in the optical afterglow, a redshift
of z=0.1685 has been identified~\citep{greiner03}.

The X-Ray Telescope (XRT) onboard \emph{Swift} began observations
of GRB~050509B 62s after the trigger of the Burst Alert telescope
(BAT)~\citep{gehrels05}. Optical and infrared data were reported
in~\citet{bloom06}. \citet{prochaska05} and \citet{bloom05}
reported a redshift of z$\sim$0.22 based on numerous absorption
features and a putative host galaxy, respectively.

GRB~050709 was discovered by HETE-2~\citep{villasenor05}. Its
prompt emission lasted 70 ms in the 3-400 keV energy band,
followed by a weaker, soft bump of $\sim$100 s duration. The
optical counterpart of this burst was observed with the Danish
1.5-m telescope at the La Silla Observatory. The observations
started 33~hours after the burst and spanned over the following 18
days~\citep{hjorth05}. Observations with the Chandra X-ray
observatory revealed a faint, uncatalogued X-ray source inside the
HETE-2 error circle~\citep{fox05}, which was coincident with a
pointlike object embedded in a bright galaxy~\citep{jensen05} at z
= 0.16~\citep{price05}.

GRB~051221A was localized by BAT~\citep{cummings05} and also
simultaneously observed by the Konus-Wind instrument. The
X-ray~\citep[$\sim10^2 - 2\times 10^6$s;][]{burrows06} and the
optical~\citep[$\sim 10^4 - 4\times10^5$s;][]{soderberg06}
afterglow light curves of GRB~051221A were well detected, while in
the radio band only one detection followed by several upper limits
are available~\citep{soderberg06}. \citet{soderberg06} detected
several bright emission lines, indicating a redshift of
$z=0.5464$.

GRB~060505 was detected by BAT in the 15-150 keV
band~\citep{palmer06,hullinger06}.  \citet{ofek06} reported the
detection of the optical transient, later confirmed by VLT FORS2
observations~\citep{thone06}.
XRT detected a source which was located about 4$\arcsec$ from a
galaxy with z=0.0894~\citep{conciatore06}.

GRB~060614 triggered both \emph{Swift}-BAT~\citep{parsons06} and
Konus-Wind~\citep{golenetskii06}. XRT found a very bright
($\sim$1300 counts s$^{-1}$) un-catalogued source inside the BAT
error circle. Ground-based optical and infrared follow-up
observations were performed using several
instruments~\citep[e.g.,][]{cobb06,schmidt06}. Based on the
detection of the host galaxy emission lines, a redshift of z = 0.125
was proposed by~\citet{price06} and confirmed by~\citet{fugazza06}.
Noted that the classification of GRB~060614 is ambiguous in the
commonly-used long/short burst scheme, since it has a long duration
but no accompanying SN~\citep{Gehrels06,Fynbo06}.

\subsection{Constraining the Model Parameters}

The available multi-frequency afterglow data are then used to
obtain the model parameters.  In this work, we have reproduced the
multi-frequency afterglow data of GRB~030329 and GRB~060614.

\begin{figure}
\begin{center}
\includegraphics[width=280pt]{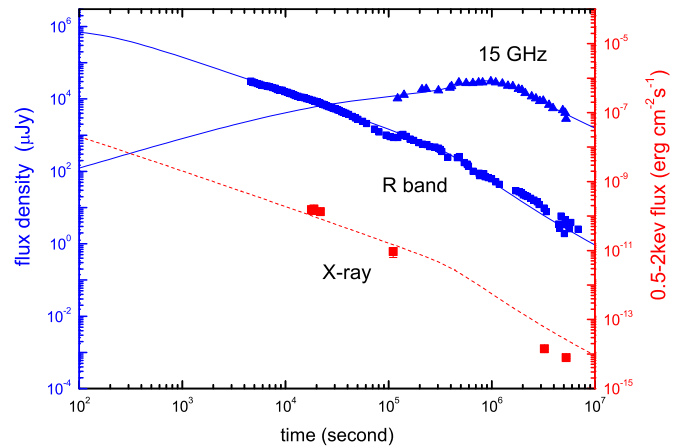}
\end{center}
\caption{GRB~030329 afterglow data in the 15~GHz~\citep{berger03},
{\em R}-band~\citep{lipkin04} and 0.5-2keV band ~\citep{tiengo04}.
Symbols indicate data points as labelled. Solid lines represent
the modeled {\em R}-band and 15 GHz emission (left plotting axis)
and the dotted line represents the modeled X-ray emission (right
plotting axis), respectively. } \label{fig:Fits}
\end{figure}

The well-sampled distinguishing afterglow behavior of GRB~030329 has
gained much attention. Some authors concentrated on the
rebrightening occurring at 1.6 days after the trigger and considered
different mechanisms to explain the rebrightening features seen in
the optical light curves~\citep{huang06}. We concentrate on the
multi-waveband emission, from radio~\citep{berger03},
optical~\citep{lipkin04} to X-ray band~\citep{tiengo04} for the
purpose in this paper. We show in Figure~\ref{fig:Fits} that, with a
certain set of parameters, the numerical results can describe the
observed data in all three wavebands.  Fluctuations were captured in
$R$ band afterglow light curves after $5\times 10^4$s from the burst
trigger, which may imply the multiple energy injection into the
outflow \citep{huang06} or a two component jet \citep{berger03}. We
ignore these details and focus on the general trend of the optical
emission (particularly, in our calculation the energy of the
relativistic ejecta is a constant). As shown in Figure
\ref{fig:Fits}, the difference is that in the time range
  $5\times 10^{4}-10^{5}$ sec our approach gives a (little bit) brighter optical
  emission, so will be the high energy emission.

\begin{figure}
\begin{center}
\includegraphics[width=280pt]{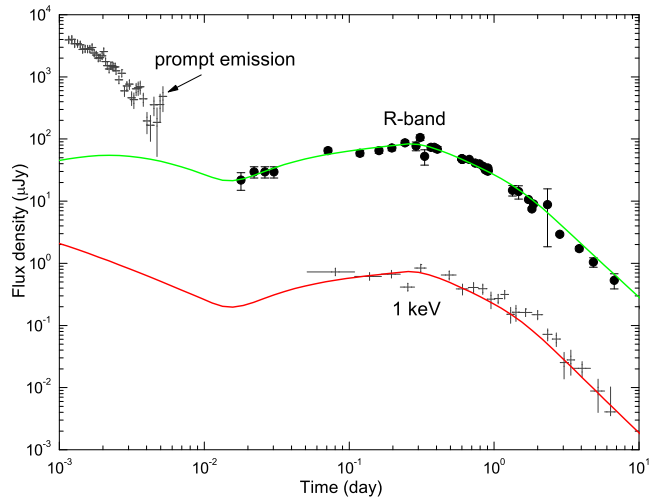}
\end{center}
\caption{GRB~060614 afterglow data in the {\em R}-band and X-ray
band~\citep[see also][]{Xu08}. Crosses represent data recorded by
\emph{Swift}-XRT (0.2-10 keV), and circles represent {\em R}-band data. Solid and dashed lines represent
the modeled {\em R}-band and 1 keV emission, respectively. }
\label{fig:Fit1}
\end{figure}

The modeled and observed afterglow light curves of GRB~060614 are
shown in Figure~\ref{fig:Fit1}. Energy
injection, starting around 30 minutes after the GRB onset, is
needed in the afterglow modeling to reproduce the increase in flux
(instead of simple power-law decay seen in other GRBs). The early
X-ray flux before 500s from the burst trigger, which is much
brighter than the modeled flux, results from the dominating
contribution from the prompt emission.

Table~\ref{sample} lists the physical parameters derived from the
afterglow modeling for GRB 030329 and GRB 060614. Parameters of
GRB~050509B, GRB~050709, GRB~051221A and GRB~060505  are taken
from the literature, also listed in Table 1.

\subsection{VHE Gamma-ray Observational Data}
We are interested in VHE observations during the afterglow phase
when the SSC is likely to dominate (see \S\ref{sect_discussion}).
VHE $\gamma$-ray afterglow data of three of the GRBs in the sample
(i.e. GRB~030329, GRB~050509B, and GRB~060505) are available.

\subsubsection{GRB~030329}
\begin{figure*}
\begin{center}
\includegraphics[width=250pt]{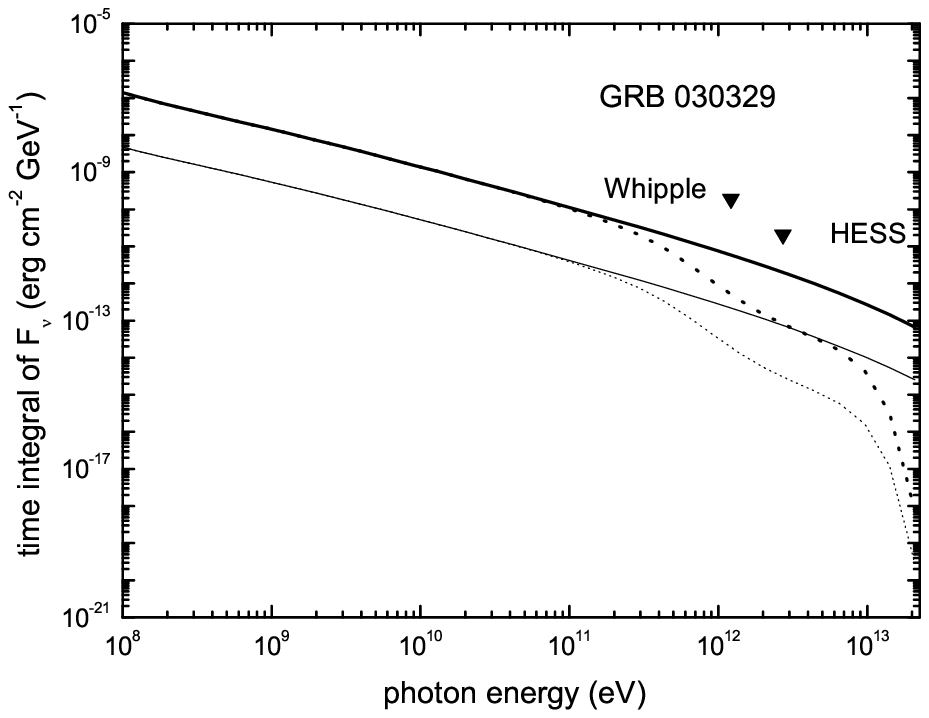}
\includegraphics[width=250pt]{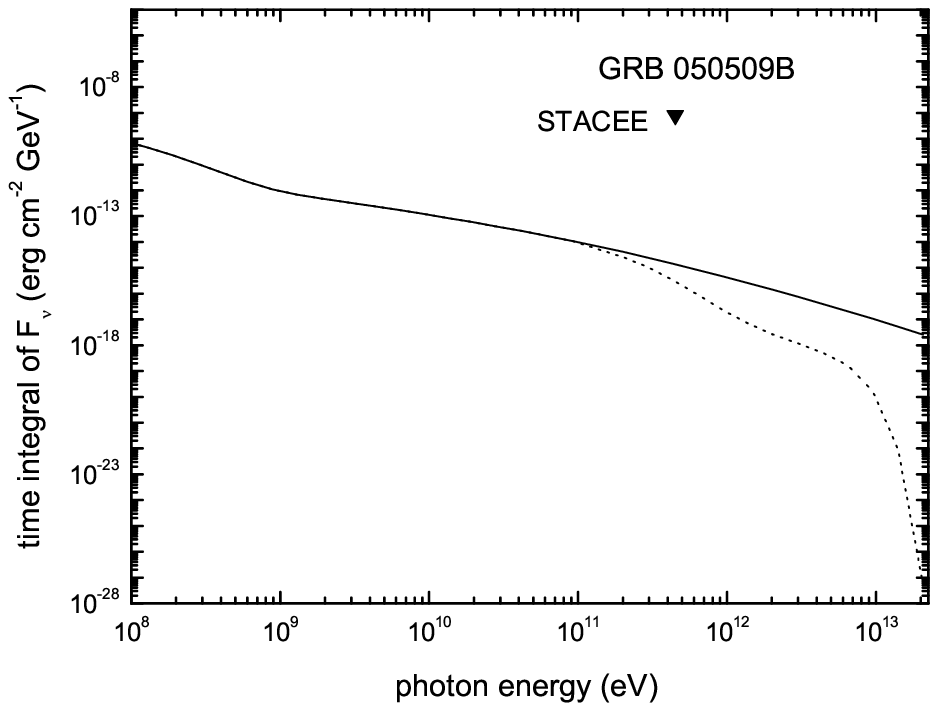}
\includegraphics[width=250pt]{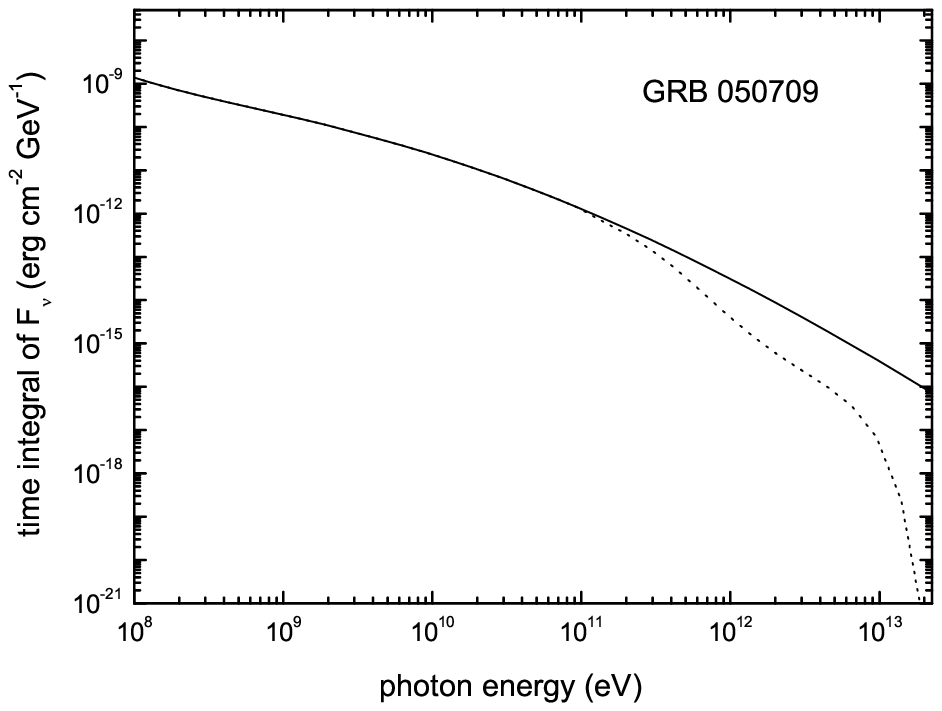}
\includegraphics[width=250pt]{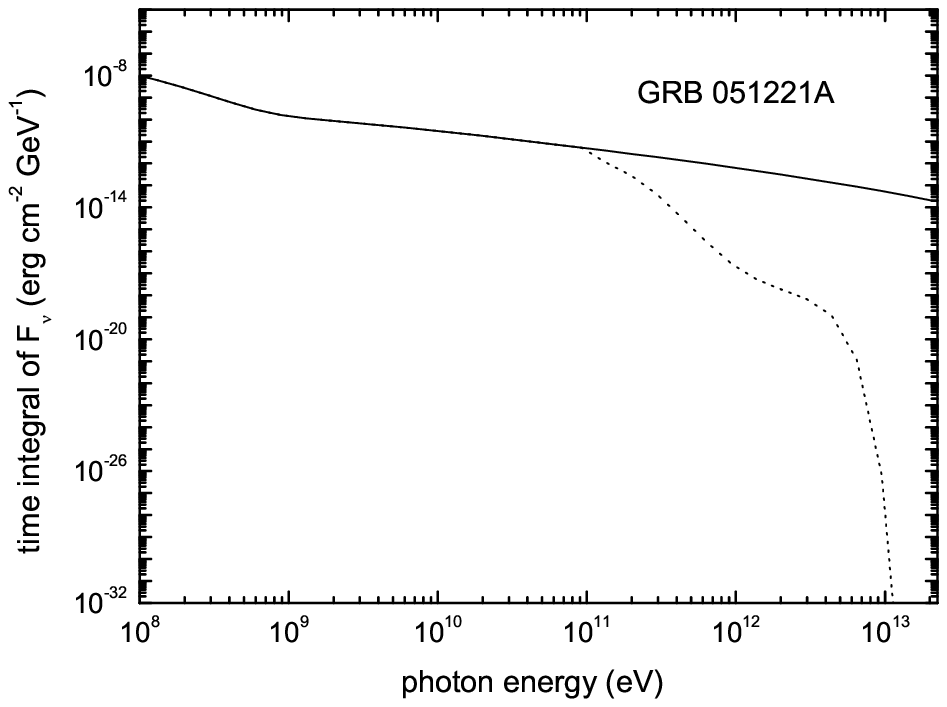}
\includegraphics[width=250pt]{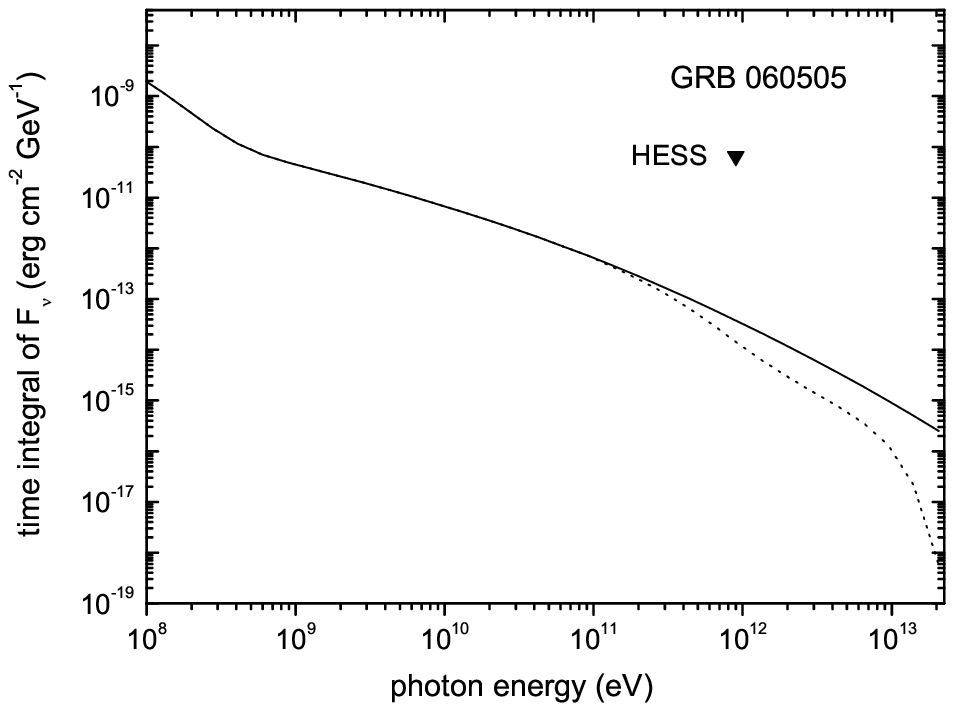}
\includegraphics[width=250pt]{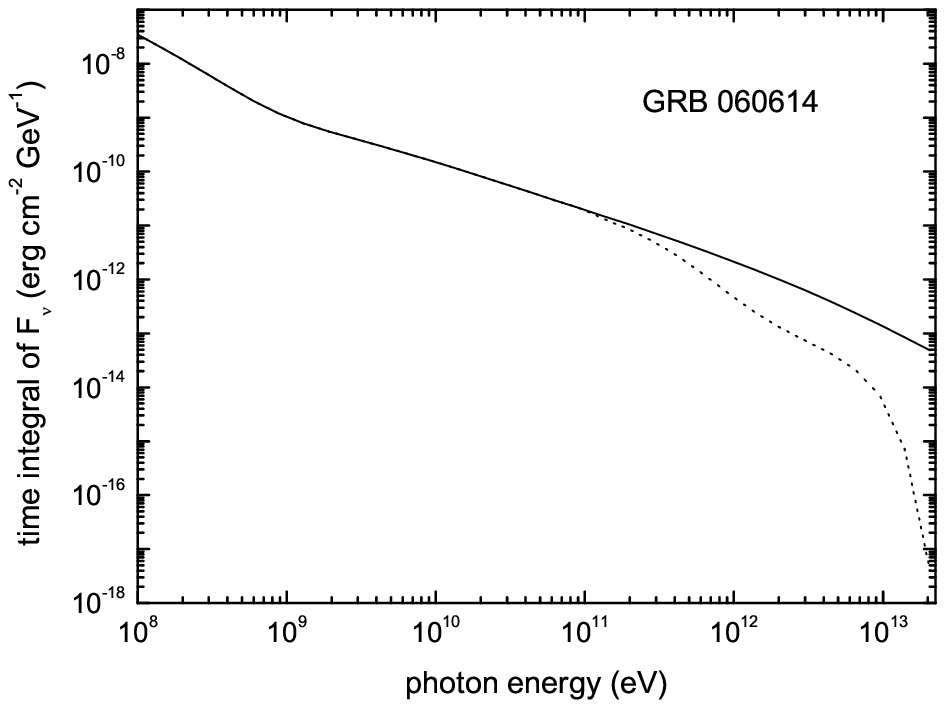}
\end{center}
\caption{Modeled time-integrated 0.1 GeV -- 20 TeV afterglow
spectra of six GRBs, in comparison with VHE upper limits
(triangles). Dotted and solid lines represent the spectra with and
without correction by the cosmic infrared background,
respectively. For GRB~030329, GRB~050509B, and GRB~060505, the
spectra were integrated over the corresponding time intervals
during which the upper limits were derived, as shown in
Table~\ref{tab:obs_time}. For GRB~030329, thick (upper) lines
indicate the modeled spectrum for the Whipple observation time,
and thin (lower) lines for the H.E.S.S. observation time. The data
points are plotted at the corresponding average photon energies.
The modeled spectra of the remaining three bursts are obtained by
integrating the spectra over a time period of 2 hours, starting
from 10~hours after the trigger.} \label{fig:HighEnergy}
\end{figure*}

\citet{horan07} reported a total of 4~hours of observations, which
spanned five nights, using the Whipple 10-m telescope. No evidence for
VHE $\gamma$-ray signal was found during any of the observation
periods. When combining all data, a flux upper limit of
$1.4\times10^{-11}\mathrm{erg}\,\mathrm{cm}^{-2}\mathrm{s}^{-1}$
was derived. The first observation, lasting for about an hour, was started 64.6~hours after the
burst. The 99.7\% c.l. flux upper
limit above an energy of $\sim400$~GeV derived from this
observation is shown in Table~\ref{flux}, as well as in
Figure~\ref{fig:HighEnergy}.

The 28-minute H.E.S.S. observation of GRB 030329 were taken 11.5
days after the burst~\citep{Tam08}. Since the burst position was
located above the northern hemisphere, the zenith angle of the GRB
observation was relatively large, i.e. 60$\degr$, resulting in an
energy threshold of 1.36~TeV. No evidence for VHE $\gamma$-ray
signal was found. The 99\% c.l. flux upper limit ($>1.36$~TeV) is
$3.4\times 10^{-11}\mathrm{erg}\,\mathrm{cm}^{-2}\mathrm{s}^{-1}$,
assuming a photon index of $\Gamma=3$.

\subsubsection{GRB~050509B}
The observations of this burst using the STACEE detector employ an
''on-off observation mode''  and contain two 28-minute on/off
pairs. The first on-source observation started 20~minutes after
the burst and the second 80~minutes after the burst. After data
quality cuts, about 18~minutes of useful on-source data remain in
each observation. No evidence for VHE $\gamma$-ray signal above
the energy threshold of 150 GeV was reported by~\citet{jarvis08}.
The 95\% c.l flux upper limits (above 150 GeV, assuming a photon
spectrum of $dN/dE \sim E^{-2.5}$) were
$3.8\times10^{-10}\mathrm{erg}\, \mathrm{cm}^{-2}\mathrm{s}^{-1}$
and $4.5\times10^{-10}\mathrm{erg}\,
\mathrm{cm}^{-2}\mathrm{s}^{-1}$ for the first and second
on-source observation, respectively (A. Jarvis private
communication).

\subsubsection{GRB~060505}
The H.E.S.S. observations began 19.4 hours after the burst and
lasted for 2 hours~\citep{Tam08}. No evidence for VHE $\gamma$-ray
signal was found. The 99\% c.l. flux upper limit ($>0.45$~TeV) is
$8.8\times 10^{-12}\mathrm{erg}\,\mathrm{cm}^{-2}\mathrm{s}^{-1}$,
assuming a photon index of $\Gamma=3$.

\subsection{Comparison to Observations}

Based on the parameters obtained in \S4.2, the GeV-TeV emission
is obtained using the code described in \S2.2.

We depict the calculated high energy afterglow spectrum in
Figure~\ref{fig:HighEnergy}, which shows the time-integrated high
energy afterglow spectrum of these six events. The solid and
dashed lines represent the intrinsic SSC spectra and corrected
spectra for each GRB, respectively. The absorption is based on the
cosmic infrared background model ``P0.45'' \citep{aharonian06a}
\footnote{This implies a gamma ray horizon at a redshift of about
0.2 (0.05) for 500 GeV (10 TeV) gamma rays.}. Such model is
constrained by the upper limits provided by two unexpectedly hard
spectra of blazars at optical/NIR wavelengths and is close to the
lower limit from integrated light of resolved galaxies.

In order to compare with the VHE observational data which are
usually given in integrated photon fluxes, we integrate the
spectra above the energy threshold. We consider first the GRBs
with VHE data. These include GRB~030329, GRB~050509B, and
GRB~060505. In Table 2 we list the modeled integrated energy
fluxes after correction due to interaction with photons from
cosmic infrared background, as well as the VHE $\gamma$-ray
observations and the derived upper limits. All predicted fluxes
are below the upper limits derived from the VHE observations.

\begin{figure}
\begin{center}
\epsfig{file=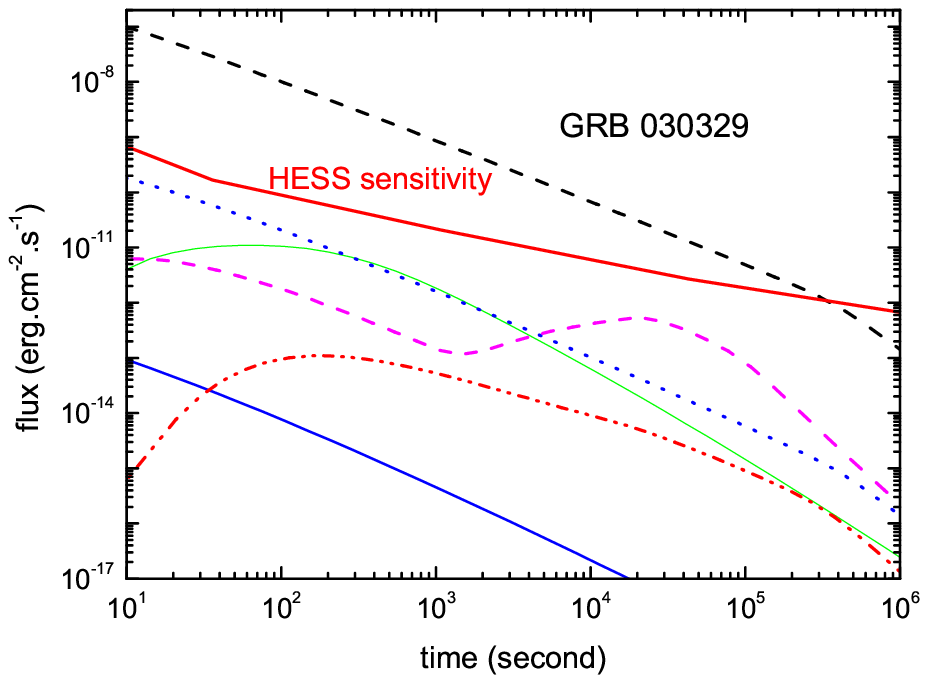,width=280pt}
\end{center}

\caption{The temporal evolution of modeled VHE integral energy
fluxes above 200~GeV for GRB 030329 (black dashed line), GRB 050509B
(blue solid line), GRB 050709 (green solid line), GRB 051221A (red
dash-dot-doted line), GRB 060505 (blue dotted line) and GRB 060614
(magenta dashed line). The red solid line represents the HESS
sensitivity if the observations start right at the GRB onset. }

\label{fig:sensitivity}
\end{figure}

We then investigate whether a sensitive VHE instrument is expected
to detect the predicted VHE signal from nearby GRBs during the late
afterglow phase. We use H.E.S.S. sensitivity as
 an example of an array of sensitive atmospheric Cherenkov
 telescopes.
The sensitivity level of H.E.S.S. detector is shown
in Figure~\ref{fig:sensitivity}, assuming a $\Gamma$=2.6
spectrum~\citep{aharonian06b}. Assuming softer spectra, the level is
higher, and the difference is about 50\% between $\Gamma$=2.0 and
$\Gamma$=3.0~\citep[c.f.][]{aharonian05}. We show the temporal
evolution of energy fluxes ($>$200GeV) of six GRBs in our sample in
Figure~\ref{fig:sensitivity}, indicating only the VHE signal from
GRB 030329 may be above the H.E.S.S. sensitivity. For GRB~030329
which is a bright burst with low redshift, the expected energy flux
would be high enough to be detected with a delayed observation time
of $\la10$ hours if GRB position was favorable, i.e. with zenith
angle $<20\degr$ (and thus an energy threshold of $\sim200$~GeV is
attained).

\section{discussion}
\label{sect_discussion}
In this paper, we have calculated the SSC emission from
the forward shock electrons following~\citet{Fan08_mn}. We shall discuss here the importance of other
radiation processes in the late afterglow phase.

Possible VHE $\gamma$-ray emission initiated from protons has been
suggested~\citep{totani98,bottcher98}. However, the
proton-synchrotron component, as well as the hadron-related
photo-meson electromagnetic components, is in most cases
overshadowed by the SSC component of electrons in the afterglow
phase. This is especially the case for the parameter values of
$\epsilon_e$ and $\epsilon_B$ used here in the modeling of these six
GRBs~\citep[e.g.,][]{zhang01b}.

Additional IC component will play a role and will enhance any VHE
emission. SSC is, in general, only a lower limit. Another possible
contribution to VHE emission is related to the ``central engine
afterglow" emission, like the flares \citep[e.g.,][]{falcone07} or
the plateaus followed by a sudden drop \citep[e.g.,][]{zhang09}. The
SSC radiation of the late internal shocks or the external inverse
Compton radiation in the external forward shock front may be able to
give rise to some VHE emission signals (see Fan \& Piran 2008 for a
review). However, no ``central engine afterglow" emission has been
reported in the late afterglow phase for the six bursts we studied.
So it is hard for us to estimate its ability to enhance the
detectability of the VHE emission.



As shown in Figure~\ref{fig:sensitivity}, we only expect detectable
signal using a sensitive ground-based $\gamma$-ray detector for a
bright and nearby GRB like GRB~030329. The rate of such nearby and
energetic GRBs is very uncertain. GRB~940217 might have been another
event of this kind~\citep{wei07}.

Several factors that reduce the chance of detecting VHE photons
can be summarized as follows: Firstly, as a result of large zenith
angles (e.g., 60$\degr$ for GRB~030329), the energy thresholds of
some observations are relatively high ($\sim$1.4~TeV). Any VHE
photon is severely attenuated by the cosmic infrared background,
unless the level is very low. Secondly, the observations were
taken at very late epochs, e.g. 11.5 days after the burst for
H.E.S.S. observations of GRB~030329, when expected VHE flux had
largely decayed. Thirdly, the fraction of low-redshift GRBs is
small, e.g. see the catalogue in \citet{Butler07}. For GRB~051221A
(at $z=0.55$) studied here, the attenuation is severe at energies
$\ga$200 GeV.

Despite these practical limitations, detection of VHE afterglow
emission of GRBs is probable. Those GRBs close enough (z$<$0.5)
and with an intrinsic high luminosity (like GRB~030329), can be
detected above $\sim$200 GeV when the observation is taken within
$\sim$10~hours after the burst. From Eq. 1, GRBs with low $z$,
large $E_0$, large  $\epsilon_e$, and small $\epsilon_B$ are more
likely to be detected in the VHE band. 
Besides, in general the
expected VHE afterglow flux decays as $t^{-\alpha}$ where $\alpha$
is at least one. Thus, observations with shorter delay are more
likely to probe the predicted VHE emission.


{\it The main purpose of this work is to provide a relatively
reliable prediction of the detectability of the VHE emission from
nearby GRBs}. It is also interesting to explore the role of the VHE
afterglow emission detection in revealing the GRB physics. Perhaps
the most robust conclusion is that the VHE emission in late
afterglow phase cannot be attributed to the synchrotron radiation of
the forward shock, for which the maximal photon energy is $\approx
30\Gamma/(1+z)~{\rm MeV}$ \citep{cw96}, where $\Gamma$ is the bulk
Lorentz factor of the decelerating outflow. In general, for
$\epsilon_B\geq \epsilon_e$, the inverse Compton emission is
expected to be very weak. So the detection of VHE afterglow emission
from GRBs requires that $\epsilon_B \ll \epsilon_e$, in support of
the current radio/optical/X-ray afterglow modeling (e.g., see
Tab.1). Together with \emph{Swift}, AGILE, FGST and some
ground-based optical telescopes, ground-based $\gamma$-ray detectors
can provide us continuous spectra in the optical to TeV energy band
during the afterglow phase. A self-consistent modeling of these data
in a very wide energy band, in principle, will impose very tight
constraint on the physical parameters and on the environment.
However, given the small number of the VHE photons expected from a
single burst, we do not expect that the VHE emission can help us a
lot to achieve such goals.

\section{conclusions}
\label{sect_conclusions} In this work, we discuss the prospect of
detecting VHE $\gamma$-rays with current ground-based detectors in
the late afterglow phase. During this phase, the dominant
radiation process in the VHE $\gamma$-ray regime is the SSC
emission from the forward shock electrons. Klein-Nishina effects
and  attenuation by the cosmic infrared background, both known to
suppress the VHE $\gamma$-ray spectra, were taken into account. To
minimize the attenuation effect, we chose a sample of six nearby
GRBs in this study. We have calculated the detailed SSC emission
numerically using the model developed by~\citet{Fan08_mn}, with a
set of parameters which are able to reproduce the available
multi-wavelength afterglow light curves. The results are
consistent with the upper limits obtained using VHE observations
of GRB~030329, GRB~050509B, and GRB~060505.
Moreover, assuming observations taken 10~hours after the burst, the
VHE signal predicted from five GRBs is below the sensitivity level
of a current sensitive atmospheric Cherenkov detectors mainly due to
the low fluence of these outflows. For those bright and nearby
bursts like GRB~030329, a VHE detection is possible even with a
delayed observation time of $\sim$10~hours.

\acknowledgments We thank the anonymous referee for  helpful
comments and D. Xu for providing us data in Figure 3. This work is
supported by the National Science Foundation (grants 10673034 and
10621303) and National Basic Research Program (973 programs
2007CB815404 and 2009CB824800) of China. YZF is also supported by a
grant from Danish National Research Foundation. PHT acknowledges
support from IMPRS-HD.

\begin{deluxetable}{lccccccccccc}
\label{sample} \tablewidth{0pt} \tablecaption{Model parameters for six nearby GRBs \label{sample}} \tablehead{ \colhead{GRB} & \colhead{z} &
\colhead{E$_0$(erg)} & \colhead{$\theta_0$} &\colhead{n(cm$^{-3}$) }
& \colhead{p} & \colhead{$\epsilon_e$} & \colhead{$\epsilon_B$} &
\colhead{L$_{eje}$} & \colhead{q} & \colhead{injection timescale(s)}
& \colhead{references}} \startdata
 030329  &0.1685 &1.4$\times10^{53}$  &0.31 &100 &2.01 &0.1 & 0.001 &\nodata &\nodata &\nodata & this work\\
 050509B &0.2248 &2.75$\times10^{48}$ &0.5  &1   &2.2  &0.15 &0.046 &\nodata &\nodata &\nodata\ &1\\
 050709  &0.16   &3.77$\times10^{50}$    &0.5  &6$\times10^{-3}$   &2.6  &0.4 &0.25 &\nodata &\nodata &\nodata &2\\
 051221A &0.5465 &$10^{52}$ &0.1  &0.01   &2.4  &0.3 &2$\times10^{-4}$
 &2$\times10^{48}$ & magnetar wind &$<1.5\times10^4$ &3\\
 060505  &0.089 &2.6$\times10^{50}$  &0.4 &1 &2.1 &0.1 & 0.008 &\nodata &\nodata &\nodata & 4\\
 060614  &0.125 &5$\times10^{50}$  &0.08 &0.05 &2.5 &0.12 & 2$\times10^{-4}$ &$10^{48}$ &0 &$10^3-2\times10^4$ & this work\\
\enddata
\tablerefs{1~\citet{bloom06}; 2~\citet{panaitescu06};
3~\citet{fan06}; 4~\citet{Xu08}}
\end{deluxetable}

\begin{deluxetable}{lccccccc}
\tablewidth{0pt} \tablecaption{VHE GRB observations and model
predictions \label{flux} } \tablehead{ \colhead{} &
\colhead{telescope} &\colhead{$\rm T_{OBS}-T_{GRB}$
\tablenotemark{a}} & \colhead{exposure} &\colhead{energy
threshold} & \colhead{energy flux upper limit set} &
\colhead{predicted energy flux} & \colhead{references}\\
\colhead{GRB} &\colhead{} &\colhead{} &\colhead{} &\colhead{(GeV)}
&\colhead{  by observations ($\rm erg\,cm^{-2}\,s^{-1}$)}
&\colhead{($\rm erg\,cm^{-2}\,s^{-1}$)} &\colhead{}} \startdata
 030329  &H.E.S.S.   &11.5 days &28 min    &1360  &$3.4\times10^{-11}$           &$8.5\times10^{-15}$  & 1\\
 030329  &Whipple &64.55 hours &65.2 min &400  &$5.8\times10^{-11}$   &$6.7\times10^{-13}$  &2\\
  050509B  &STACEE  &20 min/80 min &28 min     &150  &3.8$\times10^{-10}/4.5\times10^{-10}$           &2.2$\times10^{-16}$/5.4$\times10^{-17}$   &3\\
 060505  &H.E.S.S.   &19.4 hours &2 hours     &450    &8.8$\times10^{-12}$           &2.5$\times10^{-15}$ & 1\\
\enddata
\tablenotetext{a}{The time between the start of the GRB and the
beginning of observations for different telescopes.} \tablerefs{1
\citet{Tam08}; 2 \citet{horan07};  3 \citet{jarvis08}}
\label{tab:obs_time}
\end{deluxetable}



\epsscale{.80}
\clearpage




\end{document}